\documentclass[12pt]{article}
\usepackage{graphicx}
\usepackage{latexsym,amsmath,amsfonts,amssymb,color}
\usepackage[latin1]{inputenc}
\usepackage{cite}
\usepackage[american]{babel}
\usepackage{hyperref}
\newcommand{\be}{\begin{equation}}
\newcommand{\ee}{\end{equation}}
\newcommand{\beq}{\begin{equation}}
\newcommand{\eeq}{\end{equation}}
\newcommand{\bea}{\begin{eqnarray}}
\newcommand{\eea}{\end{eqnarray}}

\newcommand{\ba}{\begin{eqnarray}}
\newcommand{\ea}{\end{eqnarray}}
\begin{document}
\baselineskip=15.5pt
\pagestyle{plain}
\setcounter{page}{1}
%--------+---------+---------+---------+---------+---------+---------+
%Body

\def\ie{{\em i.e.},}
\def\eg{{\em e.g.},}
\newcommand{\rc}{\nonumber\\}
\newcommand{\bear}{\begin{eqnarray}}
\newcommand{\eear}{\end{eqnarray}}
\newcommand{\Tr}{\mbox{Tr}}    % trace over gauge indices
\newcommand{\ack}[1]{{\color{red}{\bf Pfft!! #1}}}

\def\a{\alpha}
\def\b{\beta}
\def\c{\gamma}
\def\d{\delta}
\def\eps{\epsilon}           % Also, \varepsilon
\def\f{\phi}               %      \varphi
\def\vf{\varphi}  \def\tvf{\hat{\varphi}}
\def\vp{\varphi}
\def\g{\gamma}
\def\h{\eta}
\def\j{\psi}
\def\k{\kappa}                    % Also, \varkappa (see below)
\def\l{\lambda}
\def\m{\mu}
\def\n{\nu}
\def\o{\omega}  \def\w{\omega}
\def\p{\pi}
\def\q{\theta}  \def\th{\theta}                  %     \vartheta
\def\r{\rho}                                     %     \varrho
\def\s{\sigma}                                   %     \varsigma
\def\t{\tau}
\def\u{\upsilon}
\def\x{\xi}
\def\z{\zeta}
\def\pt{\hat{\varphi}}
\def\tt{\hat{\theta}}
\def\lab{\label}
\def\6{\partial}
\def\wg{\wedge}
\def\atanh{{\rm arctanh}}
\def\bpsi{\bar{\psi}}
\def\bt{\bar{\theta}}
\def\bvf{\bar{\varphi}}
\def\W{\Omega}

\numberwithin{equation}{section}

\renewcommand{\theequation}{{\rm\thesection.\arabic{equation}}}

%%%%%%%%%%%%%%%%%%%%%%%%%%%%%%%%%%%%%%%%%%%%%%%%%%%%%%%%%%%%%%%%%%%%%%%%%%
%%%%%%%%%%%%%%%%        AND THERE WE GO!           %%%%%%%%%%%%%%%%%%%%%%%%
%%%%%%%%%%%%%%%%%%%%%%%%%%%%%%%%%%%%%%%%%%%%%%%%%%%%%%%%%%%%%%%%%%%%%%%%%%

\begin{flushright}
%HIP-2016-nn/TH
\end{flushright}

\begin{center}

\centerline{\Large {\bf D3-D5 theories with unquenched flavors}}

\vspace{8mm}

\renewcommand\thefootnote{\mbox{$\fnsymbol{footnote}$}}
Eduardo Conde${}^{1,2}$\footnote{econdepe@snu.ac.kr},
Hai Lin${}^{3}$\footnote{hailin@mail.tsinghua.edu.cn},
Jos\'e Manuel Pen\'\i n${}^{4,5}$\footnote{jmanpen@gmail.com},\\
Alfonso V. Ramallo${}^{4,5}$\footnote{alfonso@fpaxp1.usc.es}
and Dimitrios Zoakos${}^{6,7}$\footnote{zoakos@gmail.com}

\vspace{4mm}

${}^1${\small \sl School of Physics \& Astronomy and Center for Theoretical Physics,\\ Seoul National University, Seoul 08826, South Korea}\\
{\small \sl and} \\
 ${}^2${\small \sl Gauge, Gravity \& Strings, Center for Theoretical Physics of the Universe, \\ Institute for Basic Sciences, Daejeon 34047, South Korea}

\vskip 0.2cm
${}^3${\small \sl Yau Mathematical Sciences Center \\ Tsinghua University, Beijing, 100084, P. R. China} 
\vskip 0.2cm
${}^4${\small \sl Departamento de  F\'\i sica de Part\'\i  culas} \\
{\small \sl Universidade de Santiago de Compostela} \\
{\small \sl and} \\
${}^5${\small \sl Instituto Galego de F\'\i sica de Altas Enerx\'\i as (IGFAE)} \\
{\small \sl E-15782 Santiago de Compostela, Spain} 
\vskip 0.2cm

${}^6${\small \sl Universitat Internacional de Catalunya,\\ Immaculada 22, E-08017 Barcelona, Spain} \\
{\small \sl and} \\
${}^7${\small \sl Centro de F\'isica do Porto, Universidade do Porto, \\
Rua do Campo Alegre 687, 4169--007 Porto, Portugal
}

\end{center}

\vspace{8mm}
\numberwithin{equation}{section}
\setcounter{footnote}{0}
\renewcommand\thefootnote{\mbox{\arabic{footnote}}}

\begin{abstract}
We construct the string duals of the defect theories generated when $N_f$ flavor D5-branes intersect $N_c$ color D3-branes along a 2+1 dimensional subspace. We work in the Veneziano limit in which $N_c$ and $N_f$ are large and $N_f/N_c$ is fixed. By smearing the  D5-branes, we  find supergravity solutions that take into account the backreaction of the flavor branes and preserve two supercharges. When the flavors are massless the resulting metric displays an anisotropic Lifshitz-like scale invariance. The case of massive quarks is also considered.

\end{abstract}

\newpage
\tableofcontents

\section{Introduction}
After its original formulation \cite{Maldacena:1997re}, the AdS/CFT  holographic correspondence has been generalized and extended in many directions (see \cite{AdS_CFT_reviews} for reviews). This correspondence is a duality motivated by the twofold nature of D-branes,  as hypersurfaces  where open strings end and supergravity solutions,  which suggests that the geometry corresponding to the near-horizon limit of a stack of D-branes is related to the strong coupling limit of a gauge theory in the planar limit. An important development of the AdS/CFT duality has been the addition of flavor D-branes to the near-horizon geometry \cite{Karch:2000gx,Karch:2002sh}. These flavor branes are dual to fields transforming in the fundamental representation of the gauge group (quarks), in contrast  
with the fields in the adjoint representation which are dual to the pure supergravity solution. 

In a first approach the quarks can be considered in the quenched approximation, which amounts to neglecting the quark dynamical effects and setting the  quark fermion determinant to one. In a theory with $N_c$ colors and $N_f$ flavors the quenched approximation is well justified if $N_f<< N_c$. In the holographic dual the addition of quenched quarks corresponds to considering the flavor branes as probes in the supergravity background generated by the stack of color branes in the near-horizon limit or, equivalently, to neglecting the backreaction of the flavor branes in the geometry.  These studies of holographic quenched flavor have been very fruitful (see \cite{Erdmenger:2007cm} for a review). Indeed, by analyzing the worldvolume physics of the flavor branes, governed by the Dirac-Born-Infeld + Wess-Zumino action, a lot of physically interesting questions can be understood.

To obtain the gravity dual of a field theory with unquenched flavor one has to solve the equations of motion of supergravity with brane sources, which have Dirac $\delta$-functions with support on the location of the branes. This fact makes the problem extremely difficult to tackle. In order to bypass this difficulty one can follow the approach first proposed in \cite{Bigazzi:2005md}, which consists in considering a continuous distribution of flavor branes, in  such a way that the branes are smeared and there are no $\delta$-function sources.  Since we are substituting a discrete set of branes by a continuous distribution, this approach is only accurate if the number 
$N_f$ of flavors is large. This regime corresponds to the so-called Veneziano limit, in which both $N_c$  and $N_f$ are large and their ratio $N_c/N_f$ is fixed \cite{Veneziano:1976wm}.  The smearing technique was successfully applied to obtain geometries with flavor backreaction in several systems\cite{Casero:2006pt, Casero:2007jj, Benini:2006hh, Benini:2007gx, Bigazzi:2009bk, Conde:2011sw,Conde:2011rg}
(see \cite{Nunez:2010sf} for a detailed review and more references).  Most of these solutions are analytic. The price one has to pay for this simplification is the modification of the field theory dual. Indeed, by superposing branes with different orientations we are modifying the R-symmetry of the theory. Moreover, the smeared flavor branes are not coincident and, thus, the flavor symmetry for $N_f$ flavors is $U(1)^{N_f}$ rather than $U(N_f)$. 
Finally, the solutions with smeared branes typically preserve less supersymmetry than the ones with localized branes.

In this paper we find a geometry generated by smeared flavor D5-branes intersecting color D3-branes along a 
$(2+1)$-dimensional subspace, where the fundamental fields live.  The corresponding dual field theory is a defect theory in which $(2+1)$-dimensional matter in the fundamental representation is coupled to a gauge theory in $3+1$ dimensions. If the branes are in flat space, the dual field theory  preserves eight supercharges. The field content and lagrangian of this defect theory was determined in \cite{DeWolfe:2001pq} (see also \cite{Erdmenger:2002ex,Skenderis:2002vf}).  The mass spectra of the meson operators for the D3-D5 theory in the quenched approximation were first found in \cite{Arean:2006pk} by looking at the fluctuations of a D5-brane probe in the  $AdS_5\times {\mathbb S}^5$ background.  Actually, the D3-D5 brane system is one of the best studied examples of holography with branes. This system has a rich phase diagram, when it is considered 
at non-zero temperature,  charge density and magnetic field (see, for example \cite{Filev:2009xp,Jensen:2010ga,Evans:2010hi}). Moreover, it has been used to model the  quantum Hall effect \cite{Kristjansen:2012ny,Kristjansen:2013hma},  as a holographic model of graphene \cite{Evans:2014mva} and has appeared in the context of bubbling geometries \cite{Gomis:2006cu}. Clearly, generalizing these results to the case of unquenched flavors is of great interest. This work is a first step in this direction. 

In this paper we will consider the case in which the internal space is a general five-dimensional Sasaki-Einstein manifold and the corresponding dual field theory is a quiver theory.  In order to preserve some amount of supersymmetry at the probe level  the D5-brane must wrap a three-dimensional special Lagrangian submanifold of the Calabi-Yau cone constructed over the Sasaki-Einstein space \cite{Yamaguchi:2003ay}. This has been checked explicitly in \cite{Arean:2004mm,Canoura:2005uz,Canoura:2006es} for different internal manifolds.  To construct a backreacted solution we must include brane sources in the supergravity equations of motion. These sources induce a violation of the Bianchi identities of some of the Ramond-Ramond (RR) field strengths. In our case the presence of D5-branes implies that the RR three-form $F_3$ is not  closed anymore. Actually, the exterior derivative of $F_3$ is a four-form which encodes the RR charge distribution of the flavor branes. We will use this fact, together with the intuition obtained from the probe brane analysis, to write an ansatz for $F_3$ that could preserve some amount of the original supersymmetry. Moreover, the deformation of the metric induced by the flavor in our ansatz is similar to the one corresponding to the D3-D7 system studied in \cite{Benini:2006hh}, which consists in squashing the internal Sasaki-Einstein space in such a way that its representation as a $U(1)$ bundle over a four-dimensional K\"ahler-Einstein manifold is preserved.

It is worth recalling that exact localized supergravity solutions for the D3-D5 system have been obtained in refs. \cite{D'Hoker:2007xy,D'Hoker:2007xz}. The geometry of these solutions is of the form 
$AdS_4\times {\mathbb S}^2\times  {\mathbb S}^2 \times \Sigma$, with $\Sigma$ being a Riemann surface. These solutions  are dual to interface theories or interface operators, preserve 16 super symmetries and have $SO(2,3)\times SO(3)\times SO(3)$ isometry. Some of these geometries have topologically non-contractible three-cycles with non-vanishing three-form RR charge, which can be interpreted as due to the presence of D5-branes. However, the additional degrees of freedom corresponding to the open strings stretched between the D3 and D5 branes do not show up (they are substituted by fluxes). As argued in \cite{Assel:2011xz}, the supergravity solution by itself  should be completed by adding explicit fivebranes in the geometry. This is the point of view adopted in this paper. Indeed, our D5-branes are dynamical sources which contribute to the energy-momentum tensor and violate explicitly the Bianchi identity of the RR three-form. By smearing these sources we will be able to find simpler solutions which allow to study the effects of dynamical flavors in this class of  top-down holographic duals.

The organization of the rest of this paper is the following. In section \ref{setup}  we formulate our setup and write our ansatz for the metric and RR  forms of the type IIB theory in the case of massless flavors. We consider the case in which the D3-branes are on the tip of a cone over a general five-dimensional Sasaki-Einstein space. We also write in this section the system of first-order BPS equations for the different functions of our ansatz and we show that they can be formulated in terms of calibration forms. We end section  \ref{setup} with the demonstration that the equations of motion of supergravity plus brane sources are satisfied by any solution of the BPS system. In section \ref{Integration} we consider the integration of the BPS system. We reduce 
this first-order system to a unique second-order master equation for a new function. This master equation can be integrated  in general for the unflavored system, as shown in section \ref{unflavored_sol}. In section \ref{flavored_sol} we find a particular solution for the unflavored system, which leads to a metric displaying an anisotropic scale invariance. In section \ref{massive} we extend the ansatz to the case of massive flavors. We find the master equation for this massive case and show how to construct solutions that interpolate between the unflavored geometry in the IR and the background corresponding to massless flavors in the UV. Finally, in section \ref{conclusions} we summarize our results and discuss some lines of future research. The paper is completed with an appendix, in  which we give a detailed derivation of the BPS equations and we write the coordinate representation of two particular Sasaki-Einstein spaces.

\section{Setup and ansatz}
\label{setup}

Suppose that we have $N_c$ color  D3-branes on the tip of a cone over a  five-dimensional Sasaki-Einstein (SE) space ${\cal M}_5$ with metric $ds_{SE}^2\,=\,ds^2_{KE}\,+\,(d\tau+A)^2$,  where  $ds^2_{KE}$ is the metric of the four-dimensional
K\"ahler-Einstein (KE) base ${\cal M}_4$ and $A$ is a one-form in ${\cal M}_4$.  Moreover, we will add $N_f$ flavor D5-branes according to the array:
\beq
\label{D3D5intersection}
\begin{array}{ccccccccccl}
 &1&2&3& 4& 5&6 &7&8&9 &  \\
D3: & \times &\times &\times &\_ &\_ & \_&\_ &\_ &\_ &      \\
D5: &\times&\times&\_&\times&\times&\times&\_&\_&\_ &
\end{array}
\eeq
In (\ref{D3D5intersection})  the  directions  $4$-$9$ are those corresponding to the SE cone.  Let us consider the theory at zero temperature.  We will adopt the following ansatz for the  ten-dimensional metric in Einstein frame:
\beq
ds^2=h^{-{1\over 2}}\,\big[-(dx^0)^2+(dx^1)^2+(dx^2)^2\,+\,e^{2m}\,(dx^3)^2\big]+
h^{{1\over 2}}\,\big[dr^2\,+\,e^{2g}\,ds^2_{KE}+e^{2f}\,(d\tau+A)^2\big]\,\,,\qquad
\label{metric_ansatz}
\eeq
where $m$, $g$ and $f$ are squashing functions that depend on the radial variable $r$ and $h=h(r)$ is a warp factor. 
The type IIB supergravity background corresponding to the array written above should contain a self-dual RR five-form 
$F_5$, induced by the stack of the color D3-branes. We will adopt the following ansatz for $F_5$:
\beq
F_5\,=\,K(r)\,\big(1+*\big)\,d^4x\wedge dr\,\,,
\label{F5_ansatz}
\eeq
where $K=K(r)$ is a function to be determined. Actually, from the Bianchi identity of $F_5$ ($dF_5=0$) we can relate $K(r)$ to the functions appearing in the metric (\ref{metric_ansatz}).  We get:
\beq
K\,h^2\,e^{-m}\,e^{4g+f}\,=\,Q_c\,\,,
\label{eom_F5}
\eeq
where $Q_c$ is a constant that can be related to the number of colors $N_c$ by employing the flux quantization condition of the five-form, namely:
\beq
Q_c\,=\,{(2\pi)^4\,g_s\,\alpha'{}^{\,2}\,N_c\over {\rm Vol}({\cal M}_5)}\,\,.
\label{Qc}
\eeq

The total action of the system is the sum of the one corresponding to ten-dimensional type IIB supergravity and the action of the branes:
\beq
S\,=\,S_{IIB}\,+\,S_{branes}\,\,,
\label{total-action}
\eeq
where  $S_{branes}$ denotes the sum of the Dirac-Born-Infeld (DBI) and Wess-Zumino (WZ) actions for the flavor branes.  The $N_f$ flavor branes of our setup act as sources of the RR three-form $F_3$. Indeed, the D5-branes couple naturally to the RR six-form potential $C_{(6)}$ through the WZ term  of their worldvolume action, which is given by:
\beq
S_{WZ}\,=\,T_5\,\sum^{N_f}\,\int_{{\cal M}_6}\,\hat C_{(6)}\,\,,
\eeq
where the hat over $C_{(6)}$ denotes its pullback to the worldvolume and  $T_5$ is the tension of the D5-brane ($1/T_5\,=\,(2\pi)^5\,g_s\,(\,\alpha'\,)^3$). In the smearing approach, valid when $N_f$ is large, we substitute the discrete distribution of flavor branes by a continuous distribution with the appropriate normalization, in such a way that the smearing amounts to performing the substitution:
\beq
\sum^{N_f}\,\int_{{\cal M}_6}\,\hat C_{(6)}\,\,
\Longrightarrow\,\,
\int_{{\cal M}_{10}}\,\Xi\wedge C_{(6)}\,\,,
\eeq
where  $\Xi$  is a four-form (the smearing form) with components along the the directions orthogonal to the worldvolume of the flavor branes.  The coupling  of the flavor branes to $C_{(6)}$ modifies the Bianchi identity for $F_3$, which gets a source term proportional to $\Xi$.
In order to determine this modification, let us write the  supergravity plus branes  action (\ref{total-action})  in terms of the RR seven-form $F_{(7)}$ and its six-form potential $C_{(6)}$. This action contains a contribution of the form:
\beq
-{1\over 2\kappa_{10}^2}\,\,{1\over 2}\,\,\int_{{\cal M}_{10}}\,e^{-\phi}\,\,F_{(7)}\wedge *F_{(7)}\,+\,T_{5}\,\int_{{\cal M}_{10}}\,C_{(6)}\wedge \Xi\,\,,
\label{C6-action}
\eeq
where  $2\,{\kappa}_{10}^2\,=\,(2\pi)^7\,g_s^2\,(\,\alpha'\,)^4$. 
The equation of motion of $C_{(6)}$ derived from (\ref{C6-action})  gives rise to the Maxwell equation for $F_{(7)}$ with $\Xi$ playing the role of a source, which is just:
\beq
d \Big(e^{-\phi}\,*F_{(7)}\,\Big)\,=\,-2\kappa_{10}^2\,T_{5}\,\,\Xi\,\,.
\label{Maxwell-F7}
\eeq
Taking into account that $F_{(3)}=-e^{-\phi}\,*\,F_{(7)}$, we get  that (\ref{Maxwell-F7}) is equivalent to
 the following violation of Bianchi identity of  $F_{(3)}$:
\beq
dF_3\,=\,2\,\kappa_{10}^2\,T_5\,\Xi\,\,.
\eeq
The four-form $\Xi$ is just the RR charge distribution due to the presence of the D5-branes. Clearly, $\Xi$ is non-zero on the location of the sources. In a localized setup, in which the $N_f$ branes are on top of each other,  $\Xi$ will contain Dirac $\delta$-functions and finding the corresponding backreacted geometry is technically a very complicated task  although, as discussed above, supergravity solutions for the D3-D5 intersection have indeed been found \cite{D'Hoker:2007xy,D'Hoker:2007xz,Assel:2011xz}.  Here we avoid this difficulty by separating  the $N_f$ branes and distributing them homogeneously along the internal manifold in such a way that, in the limit in which $N_f$ is large, they can be described by a  continuous  charge distribution $\Xi$.  

Instead of trying to specify explicitly  the family of flavor branes of our setup, let us formulate directly an ansatz for the $F_3$ leading to a smearing form compatible with the preservation of some amount of supersymmetry.  We will assume that our flavors are massless, which implies that the flavor branes reach the origin $r=0$ and that the smearing form $\Xi$ is homogeneous in $r$, \ie\ independent of the radial coordinate. From our array we notice that $x^3$ is a direction orthogonal to the D5-branes. Therefore, one of the legs of the four-form $\Xi$ (and of $F_3$) should be along the $x^3$ direction, whereas the others should lie along the internal space. In order to specify this internal structure,  let $\{e^i\}$ ($i=1, \cdots, 4$) be a canonical basis of vielbein one-forms for the KE space ($ds^2_{KE}=\sum_i (e^i)^2$). In this basis the K\"ahler two-form $J_{KE}$ of ${\cal M}_4$ can be written simply as:
\beq
J_{KE}\,=\,e^1\wedge e^2+e^3\wedge e^4\,\,,
\eeq
and it is related to the one-form $A$ in (\ref{metric_ansatz})  as:
\beq
J_{KE}\,=\,{dA\over 2}\,\,.
\label{KE_pot}
\eeq
The explicit coordinate form of  the $e^i$'s and $A$ for the cases in which 
${\cal M}_5=T^{1,1}, {\mathbb S}^5$ is given in the appendix. 
Let us next introduce the complex two-form $\Omega_2$ as:
\beq
\Omega_2\,=\,(e^1+i e^2)\wedge (e^3+i e^4)\,\,.
\eeq
This form satisfies:
\beq
d\Omega_2\,=\,3\,i\,\Omega_2\,\wedge A\,\,.
\label{d_Omega2}
\eeq
Therefore, if we define $\hat\Omega_2$ as:
\beq
\hat\Omega_2\,=\,e^{3 i\tau}\,\Omega_2\,\,,
\eeq
it follows from (\ref{d_Omega2}) that the exterior derivative of $\hat\Omega_2$ is given by:\footnote{
The two-form $\hat\Omega_2$ is related to the holomorphic  $(3,0)$-form of the Calabi-Yau cone with metric
$ds^2_{CY}\,=\,dr^2\,+\,r^2\,ds_{SE}^2$ as:
\beq
\Omega_{CY}\,=\,r^2\,\hat \Omega_2\,\wedge (dr+ir\,(d\tau+A))\,\,.
\eeq
The closure of $\Omega_{CY}$ implies (\ref{d_Omega2}).}
\beq
d\hat\Omega_2\,=\,3\,i\,\hat\Omega_2\,\wedge (d\tau+A)\,\,.
\label{d_hat_Omega2}
\eeq
Let us now separate real and imaginary parts of $\hat \Omega_2$.  From (\ref{d_hat_Omega2}) we obtain:
\beq
d \,{\rm Im}\,\hat \Omega_2\,=\,3\,{\rm Re}\,\hat \Omega_2\,\wedge (d\tau+A)\,\,.
\label{d_Im_hat_Omega_2}
\eeq
Let us now write our ansatz for $F_3$ as:
\beq
F_3\,=\,Q_f\,dx^3\wedge {\rm Im }\,\hat\Omega_2\,\,,
\label{F3_ansatz}
\eeq
where $Q_f$ is  a constant  proportional to the number of flavors $N_f$. More explicitly, $F_3$ can be written as:
\beq
F_3\,=\,Q_f\,dx^3\wedge \Big[e^1\wedge (\cos (3\tau)\,e^4+\sin(3\tau)\, e^3\,)\,+\,
e^2\wedge (\cos (3\tau)\,e^3-\sin(3\tau)\, e^4\,)\Big]\,\,.
\label{F3_ansatz_explicit}
\eeq
It follows from (\ref{d_Im_hat_Omega_2}) that the modified Bianchi identity for $F_3$ is:
\beq
dF_3\,=\,-3 \,Q_f\,dx^3\wedge {\rm Re}\,\hat\Omega_2\wedge (d\tau+A)\,\,.
\label{dF_3}
\eeq
The smearing form $\Xi$ can be read from the right-hand side of (\ref{dF_3}). Notice that $\Xi$ does not depend on $x^3$ (it only depends on $dx^3$), which means that we are homogeneously distributing  our flavor branes in the $x^3$ direction. 

To find a supersymmetric solution for our ansatz for the metric, dilaton and RR forms, we have to consider the supersymmetric variations of type IIB supergravity and find the corresponding first-order BPS equations ensuring the existence of Killing spinors. The detailed analysis of these SUSY variations is performed in the appendix. The resulting BPS system for the different  functions of the ansatz is:
\bear
&&h'\,=\,-Q_c\,e^{-4g-f}\,-\,Q_f\,e^{{\phi\over 2}-m-2g}\,h\,\,,\rc\rc
&&\phi'\,=\,Q_f\,e^{{\phi\over 2}\,-\,m\,-2g}\,\,,\rc\rc
&&m'\,=-\,Q_f\,e^{{\phi\over 2}\,-\,m\,-2g}\,\,,\rc\rc
&&g'\,=\,e^{f-2g}\,\,,\rc\rc
&&f'\,=\,3\,e^{-f}\,-\,2e^{f-2g}\,+\,{Q_f\over 2}\,e^{{\phi\over 2}-m-2g}\,\,.
\label{BPS_system}
\eear
In the vielbein basis (\ref{vielbein}),  the Killing spinor takes the form:
\beq
\epsilon\,=\,h^{-{1\over 8}}\,e^{{3\over 2}\,i\sigma_2\,\tau}\,\eta\,\,,
\label{Killing_spinor}
\eeq
where $\sigma_2$ is a Pauli matrix and $\eta$ is a doublet of constant Majorana-Weyl spinors, characterized by the four projections in  (\ref{D5_projection}) and (\ref{CY_proyection}). Therefore, the solutions of (\ref{BPS_system})  give rise to a  ten-dimensional supersymmetric background which preserves two supercharges.

Interestingly, one can write the BPS equations in terms of generalized calibration forms. As we are dealing with a system with two types of branes, we expect to have two of such calibration forms. To begin with we will have a four-form ${\cal K}_{(4)}$ that will calibrate the geometry of the D3-branes within our background.  This form can be defined as a fermion bilinear constructed from the Killing spinor $\eta$ as:
\beq
{\cal K}_{(4)}\,=\,{1\over 4!}\,{{\cal K}_{(4)}}_{a_1\cdots a_4}\,E^{a_1\,\cdots a_4}\,\,,
\qquad \qquad 
{{\cal K}_{(4)}}_{a_1\cdots a_4}\,=\,\eta^{\dagger}\,i\s_2\,\Gamma_{a_1\,\cdots a_4}\,\eta\,\,,
\eeq
where $E^{a_1\,a_2\,\cdots}$ denotes $E^{a_1}\wedge E^{a_2}\wedge\cdots$, with the $E^a$'s being the one-forms of our 10d vielbein basis. From the projections imposed on our Killing spinors, we get that ${\cal K}_{(4)}$ is given by:
\beq
{\cal K}_{(4)}\,=\,E^{x^0\,x^1\,x^2\,x^3}\,\,,
\eeq
which is quite natural given the fact that our D3-branes wrap the Minkowski part of the 10d spacetime. Moreover, we should also have a six-form ${\cal K}_{(6)}$ calibrating the worldvolume of the D5-branes. In terms of fermion bilinears, ${\cal K}_{(6)}$ is defined as:
\beq
{\cal K}_{(6)}\,=\,{1\over 6!}\,{{\cal K}_{(6)}}_{a_1\cdots a_6}\,E^{a_1\,\cdots a_6}\,\,,
\qquad \qquad 
{{\cal K}_{(6)}}_{a_1\cdots a_6}\,=\,\,\eta^{\dagger}\,\s_1\,\Gamma_{a_1\,\cdots a_6}\,\eta\,\,.
\label{K-bilinear}
\eeq
The explicit form of ${\cal K}_{(6)}$ can also be determined from the projections satisfied by the spinor $\eta$. Our D5-branes are extended along three Minkowski and three internal directions. We expect  the G-structure of our geometry to play a key role in determining the components of  ${\cal K}_{(6)}$ along the the internal manifold. We are dealing here with an SU(3) structure generated by D5-branes wrapping a three-cycle. This SU(3) structure is endowed with a 
K\"ahler two-form $J$ and a holomorphic three-form $\Omega$, given by:
\bear
&&J\,=\,E^1\wedge E^2\,+\,E^3\wedge E^4\,+\,E^r\wedge E^5\,\,,\rc\rc
&&\Omega\,=\,e^{3i\tau}\,\big(E^1+iE^2\big)\wedge \big(E^3+iE^4\big)\wedge \big(E^r+iE^5\big)\,\,.
\eear
It can be checked from the projections satisfied by the spinor $\eta$ that ${\cal K}_{(6)}$ can be written in terms of the
real part of $\Omega$ as:
\beq
{\cal K}_{(6)}\,=\,E^{x^0\,x^1\,x^2}\,\wedge \,{\rm Re}\,\Omega\,\,.
\label{K_6_expression}
\eeq
Moreover we can write the full set of SUSY-preserving conditions (\ref{BPS_system})   as:
\bear
&&d\,{\cal K}_{(4)}+*\,d{\cal K}_{(4)}\,=\,F_5\,,\rc\rc
&&e^{-\phi}\,d\,\big(e^{{\phi\over 2}}\,{\cal K}_{(6)}\big)\,=\,*\,F_3\,\,,\rc\rc
&&d\big(e^{{\phi\over 2}}\,h^{-{1\over 2}}\,*{\cal K}_{(6)}\big)\,=\,0\,\,,\rc\rc
&&d\,\big(h^{-{1\over 2}}\,J\big)\,=\,0\,\,,\rc\rc
&&d\,\big(e^{\phi+m}\big)\,=\,0.
\label{BPS_system_forms}
\eear
Notice that (\ref{BPS_system_forms}) agrees with the classification obtained in \cite{Gauntlett:2003cy}. 
As a non-trivial check of our BPS system of  first-order differential equations, let us verify that the second-order equations of motion of the  supergravity plus branes system are satisfied if (\ref{Killing_spinor}) holds. These equations of motion follow from the action (\ref{total-action}), which we now write explicitly.  The action of type IIB supergravity in Einstein frame is:
\beq
S_{IIB} = \frac{1}{2\kappa_{10}^2} \left[ \int d^{10}x\, \sqrt{-g} \left( R - \frac{1}{2} \partial_{\mu} \phi \partial^{\mu} \phi \right) -  \int \left( \frac{1}{2}\,e^{\phi} F_{3} \wedge *F_{3} \,+\,{1\over 4}\,
F_{5} \wedge *F_{5} \right) \right]\,\,,
\eeq
whereas the action of our calibrated distribution of flavor branes reads
\beq
	S_{\text{branes}} = - T_5 \int \Big(e^{\phi/2} {\cal K}_{(6)} - C_{(6)}  \Big) \wedge \Xi\,\,,
\eeq
where ${\cal K}_{(6)}$ is the calibration six-form written in (\ref{K_6_expression}) and  the smearing form 
$\Xi$ can be read from the right-hand side of (\ref{dF_3}). In terms of our vielbein basis, $\Xi$ can be written as:
\beq
\kappa_{10}^2\,T_5\,\Xi\,=\,
-{3\over 2}\,Q_f\, h^{-{1\over 2}}\,e^{-f-2g-m}\,E^{x^3}\wedge {\rm Re}
\Big[ e^{3i\tau}\, (E^1+iE^2)\wedge (E^3+iE^4)\wedge E^5\Big]\,\,.
\eeq
The equation of motion of the RR flux $F_5$ is $dF_5=0$ and it is satisfied by our ansatz if (\ref{eom_F5}) is fulfilled. Moreover,  the equation of motion for the three-form flux $F_3$ reads:
\beq
d \left( e^{\phi} * F_{3} \right) = 0\,\,.
\label{eom_F3}
\eeq
and it is satisfied if the BPS system holds since, according to the second equation in (\ref{BPS_system_forms}), 
$e^{\phi}\,* F_{3}$ is an exact seven-form.  In order to write the the dilaton and the Einstein equations, we define first the following notation:
\beq
	\o_{(p)} \lrcorner \l_{(p)} = \frac{1}{p!} \o^{\mu_1 ... \mu_p} \l_{\mu_1 ... \mu_p}\,\,,
\eeq
for any two $p$-forms $\omega_{(p)}$ and $\lambda_{(p)}$. One can easily prove that, in a ten-dimensional manifold, one has:
\beq
	\int \o_{(p)} \wedge \l_{(10-p)} = - \int d^{10}x\, \sqrt{-g} \l \lrcorner (*\o)\,\,.
\eeq
Using these results, we can write the equation of motion  of the dilaton as:
\beq
	\frac{1}{\sqrt{-g}} \partial_{\mu} \left( \sqrt{-g} g^{\mu \nu} \partial_{\nu} \phi \right) = \frac{1}{12} 
	e^{\phi} F_{3}^2 -  \kappa_{10}^2\,T_5\, e^{{\phi\over 2}}\,\, \Xi \lrcorner (* {\cal K}_{(6)})\,\,,
	\label{eom_dilaton}
\eeq
and one can check that  is satisfied when the functions of our ansatz are solutions of the system (\ref{BPS_system}). 
Let us now write the  Einstein equations that follow from the action (\ref{total-action}) as:
\bear
&&R_{\mu\nu}-{1\over 2}\,g_{\mu\nu}\,R\,=\,
{1\over 2}\,\big(\partial_{\mu}\phi\,\partial_{\nu}\phi-
{1\over 2}\,g_{\mu\nu}\,\partial_{\lambda}\phi\,\partial^{\lambda}\phi\big)+
{1\over 2 \cdot 3!}\,\big(3
{F_{(3)}}_{\mu\rho\sigma}\,{F_{(3)}}_{\nu}^{\,\,\,\rho\sigma}
-{1\over 2}\,g_{\mu\nu}\,{F_{(3)}}_{\lambda\rho\sigma}\,{F_{(3)}}^{\lambda\rho\sigma}\big)\,+\,\rc\rc
&&\qquad\qquad\qquad\qquad
+{1\over 4 \cdot 5!}\,\big(5
{F_{(5)}}_{\mu\rho\sigma\alpha\beta}\,{F_{(5)}}_{\nu}^{\,\,\,\rho\sigma\alpha\beta}
-{1\over 2}\,g_{\mu\nu}\,{F_{(5)}}_{\lambda\rho\sigma\alpha\beta}\,
{F_{(5)}}^{\lambda\rho\sigma\alpha\beta}\big)\,+\,T^{flav}_{\mu\nu}\,\,,
\label{Einstein_eq}
\eear
where $T_{\m\n}^{\text{flav}}$ is the energy-momentum tensor coming from the  action of the flavor branes. It is given by:
\beq
	T_{\m\n}^{\text{flav}} = \kappa_{10}^2\,T_5\, e^{{\phi\over 2}} \Big[  g_{\m\n}\,\Xi \lrcorner (*{\cal K}_{(6)}) -
	{1\over 3!}\,
	 \Xi_{\m \r_1 \r_2 \r_3} \left(* {\cal K}_{(6)} \right)_{\n}^{\phantom{\n} \r_1 \r_2 \r_3} \Big]\,\,.
	\label{Tmu-nu}
\eeq
It follows from (\ref{Tmu-nu}) that
the non-zero components of $T_{\m\n}^{\text{flav}}$, in flat coordinates, are given by:
\bear
&&T_{x^{\mu}\,x^{\nu}}^{\text{flav}}\,=\,-\,3\,Q_f\,h^{-{1\over 2}}\,e^{{\phi\over 2}-f-2g-m}\,\eta_{\mu\nu}\,\,,
\qquad\qquad
(\mu,\nu=0,1,2)\,\,,\rc\rc
&&T_{rr}^{\text{flav}}\,=\,-\,3\,Q_f\,h^{-{1\over 2}}\,e^{{\phi\over 2}-f-2g-m}\,\,,\rc\rc
&&T_{ij}^{\text{flav}}\,=\,-\,{3\,Q_f\over 2}\,h^{-{1\over 2}}\,e^{{\phi\over 2}-f-2g-m}\,\delta_{ij}\,\,,
\qquad\qquad
(i,j=1,\cdots, 4)\,\,,
\label{Tmu_nu_massless_components}
\eear
and one can readily demonstrate that  (\ref{Einstein_eq}) holds along the different directions.

\section{Integration of the BPS system}
\label{Integration}

Let us now consider  the integration of the first-order system (\ref{BPS_system}). Clearly, the warp factor $h$ only appears in the first equation in (\ref{BPS_system}), which can be integrated once the other functions are known.  Moreover, since $\phi'=-m'$ in (\ref{BPS_system}),  we can take without loss of generality:
\beq
\phi\,=\,-m\,\,,
\label{phi_m}
\eeq
and, thus, we are left with three equations for $\phi'$, $g'$ and $f'$.  We can also rewrite the function $K(r)$ appearing in the ansatz (\ref{F5_ansatz}) for $F_5$ in a very convenient way. First, we notice that using (\ref{phi_m}) the first equation in (\ref{BPS_system}) can be written as:
\beq
e^{-4g-f}\,Q_c\,=\,-h'\,-\,\phi'\,h\,\,.
\eeq
Plugging this result in (\ref{eom_F5}) we arrive at the following expression of $K(r)$:
\beq
K(r)\,=\,-h^{-2}\,e^{-\phi}\,\big(h'\,+\,\phi'\,h\big)\,\,,
\eeq
which can be simply recast as:
\beq
K(r)\,=\,\partial_r\,\big(e^{-\phi}\,h^{-1}\big)\,\,.
\eeq
Thus, the RR five-form $F_5$ for our solutions can be written as:
\beq
F_5\,=\,\partial_r\,\big(e^{-\phi}\,h^{-1}\big)\,\big(1+*\big)\,d^4x\wedge dr\,\,.
\label{F5_sol}
\eeq

In order to continue with the integration of the BPS equations, let us next 
introduce a new radial variable $\r$ related to the old one as:
\beq
\frac{d\r}{dr}=e^{-2g}\,.
\label{rho_coordinate_def}
\eeq
Denoting with a dot the  differentiation with respect to $\r$, 
the BPS system reduces to:
\beq\begin{aligned}
\dot{\f}\,&=\,Q_f\,e^{\frac{3\f}{2}}\,,\\
\dot{g}\,&=\,e^{f}\,,\\
\dot{f}\,&=\,3\,e^{-f+2g}-2\,e^{f}\,+\,{Q_f\over 2}\,e^{\frac{3\f}{2}}\,.
\label{eqn:BPSdots}
\end{aligned}\eeq
Let us integrate first the equation of the dilaton in (\ref{eqn:BPSdots}). Let $\phi_0$ be the value of the dilaton at $\rho=0$ and let us define the flavor deformation parameter $\varepsilon$ as:
\beq
\varepsilon\,=\,{3\over 2}\,Q_f\,e^{{3\over 2}\,\phi_0}\,\,.
\label{epsilon_def}
\eeq
Then, as a function of $\rho$, the dilaton can be written as:
\beq
e^{{3\over 2}\,\phi}\,=\,{e^{{3\over 2}\,\phi_0}\over 1-\varepsilon\,\rho}\,\,.
\label{dilaton_rho}
\eeq
Since the left-hand side of (\ref{dilaton_rho}) cannot be negative, it follows that 
the allowed  range of the  $\rho$  variable is  $-\infty<\rho<\varepsilon^{-1}$ (the IR corresponds to $\rho\to-\infty$, while the UV is the region $\rho\to\epsilon^{-1}$). 
Notice  that $\phi$ grows when $\rho$ is increased and that $\phi$ diverges at the UV endpoint $\rho=\varepsilon^{-1}$.  Using (\ref{dilaton_rho}) the remaining 
equations for $g$ and $f$ are:
\bear
&&\dot{g}\,=\,e^{f}\,\,,\rc\rc
&&\dot{f}\,=\,3\,e^{-f+2g}-2\,e^{f}\,+\,{1\over 3}\,{\varepsilon\over 1-\varepsilon\,\rho}\,\,.
\eear
One can readily show that this system of two first-order equations is equivalent to the following second-order equation for $g$:
\beq
\ddot{g}\,+\,2(\dot{g})^2\,-\,3e^{2g}\,-\,{1\over 3}\,{\varepsilon\over 1-\varepsilon\,\rho}\,\dot g\,=\,0\,\,.
\label{g_master_eq}
\eeq
Let us next  rewrite (\ref{g_master_eq}) in terms of the new function $G$, defined as:
\beq
G\,\equiv\,e^{2g}\,\,.
\eeq
We arrive at the following non-linear master equation:
\beq
\ddot G\,-\,6\,G^2\,-\,{1\over 3}\,{\varepsilon\over 1-\varepsilon \rho}\,\dot G\,=\,0\,\,.
\label{master_eq}
\eeq
Notice that, if we know $G$, we can easily get the two functions $f$ and $g$ as:
\beq
e^{g}\,=\,\sqrt{G}\,\,,\qquad\qquad
e^{f}\,=\,{\dot G\over 2G}\,\,.
\label{f_g_G}
\eeq

\subsection{The unflavored solution}
\label{unflavored_sol}

Let us consider for a while the unflavored case $\varepsilon\,=\,0$. In this case the dilaton is constant and 
the master equation (\ref{master_eq})  becomes:
\beq
\ddot G_{0}\,-\,6\,G_{0}^2\,=\,0\,\,,
\label{master_eq_unflavored}
\eeq
where $G_{0}(\rho)\equiv G(\rho)_{\varepsilon=0}$.  By multiplying (\ref{master_eq_unflavored})  by $\dot G_{0}$ one immediately realizes that this equation can be integrated once as:
\beq
\dot G_0^2\,=\,4\,G_0^3\,-\,g_3\,\,,
\label{first_integral_unflavored}
\eeq
where $g_3$ is an integration constant. The first-order differential equation (\ref{first_integral_unflavored}) is  well-known. Indeed,  the Weierstrass function $\wp(\rho;g_2, g_3)$ satisfies the differential equation:
\beq
\dot\wp^2\,=\,4\,\wp^3\,-\,g_2\,\wp\,-\,g_3\,\,,
\label{wp-dif-eq}
\eeq
where $g_2$ and $g_3$ are the so-called lattice invariants. Clearly, our first integral (\ref{first_integral_unflavored}) is just (\ref{wp-dif-eq})  for  $g_2=0$. Therefore, up to a constant shift in the $\rho$ coordinate, we can write the general solution of (\ref{master_eq_unflavored}) as:
\beq
G_0(\rho)\,=\,\wp(\rho;0, g_3)\,\,.
\label{general_unflavor_G}
\eeq
In order to write neatly the metric corresponding to the solution (\ref{general_unflavor_G}),  we change to a new radial variable $\zeta$, defined as:
\beq
\zeta\,=\,\sqrt{G_0}\,=\,e^{g}\,\,.
\eeq
Then, one can show that:
\beq
e^{f}\,=\,\zeta\,\sqrt{k(\zeta)}\,\,,
\qquad\qquad
dr\,=\,{d\zeta\over \sqrt{k(\zeta)}}\,\,,
\eeq
where $k(\zeta)$ is defined as:
\beq
k(\zeta)\,\equiv\,1-{b^6\over \zeta^6}\,\,,
\eeq
with $b^6=g_3/4$. Then, the unflavored  ten-dimensional metric takes the form:
\beq
ds^2_{unflav}\,=\,h^{-{1\over 2}}\,dx_{1,3}^2\,+\,h^{{1\over 2}}\,
\Big[{d\zeta^2\over k(\zeta)}\,+\,\zeta^2\,ds^2_{KE}
\,+\,\zeta^2\, k(\zeta)\,\,(d\tau+A)^2\Big]\,\,,
\label{gen_unflavored_metric}
\eeq
where the warp factor $h$ can be determined by integrating the first equation in (\ref{BPS_system}) (see, for example appendix B in \cite{Benini:2006hh} for its explicit expression).  The geometry (\ref{gen_unflavored_metric}) for $b$ real and positive describes $N_c$ smeared D3-branes on the blown-up 4-cycle of the Calabi-Yau \cite{PandoZayas:2001iw, Benvenuti:2005qb}. Indeed, in the six-dimensional part of the metric (\ref{gen_unflavored_metric})  the
K\"ahler-Einstein cycle is blown up at $\zeta=b$.  The gauge theory dual to this local K\"ahler deformation of the Calabi-Yau cone is a deformation of the superconformal theory due to the insertion of a VEV of a dimension 6 operator \cite{Benvenuti:2005qb}. By choosing appropriately the integration constant in the warp factor $h$ one can show that the solution becomes asymptotically  $AdS_5\times {\cal M}_5$  for large values of the radial variable $\zeta$. In the particular case with $b=g_3=0$ we recover the conformal $AdS_5\times {\cal M}_5$ solution. Interestingly, in this case the master function $G_0(\rho)$ is simply:
\beq
G_0(\rho)_{g_3=0}\,=\,\wp(\rho;0, 0)\,=\,{1\over \rho^2}\,\,.
\label{G_0_scaling}
\eeq

\subsection{A flavored solution}
\label{flavored_sol}

Let us now come back to the $\varepsilon\not=0$ model. The master equation (\ref{master_eq}) can be regarded a deformation of the unflavored one. We have not been able to solve analytically (\ref{master_eq})  in general. However, is rather easy to find a simple analytic solution similar to the scaling one in (\ref{G_0_scaling}).  This solution is simply:
\beq
G\,=\,{8\over 9}\,{1\over (\varepsilon^{-1}\,-\,\rho)^2}\,\,.
\eeq
By using (\ref{f_g_G}) we can straightforwardly obtain the functions $f$ and $g$ as:
\beq
e^{g}\,=\,{2\sqrt{2}\over 3}\,{1\over \varepsilon^{-1}-\rho}
\,\,,
\qquad\qquad
e^{f}\,=\,{1\over \varepsilon^{-1}-\rho}\,\,,
\eeq
where we have taken into account that $-\infty < \rho<\varepsilon^{-1}$. 
Let us write this solution in terms of the original variable $r$, which is related  to $\rho$ as:
\beq
r\,=\,\int\,G(\rho)\,d\rho\,=\,{8\over 9}\,{1\over \varepsilon^{-1}-\rho}\,\,,
\eeq
where we have adjusted the integration constant in such a way that the radial variable $r$ takes values in the range $0\le\,r\,<\infty$. In terms of $r$,  the dilaton is:
\beq
e^{{3\phi\over 2}}\,=\,{3\over 4\,Q_f}\,r\,\,,
\eeq
where we have used (\ref{epsilon_def}). The functions $f$ and $g$ are:
\beq
e^{g}\,=\,{{3\over 2\sqrt{2}}}\,r\,\,,
\qquad\qquad
e^{f}\,=\,{9\over 8}\,r\,\,.
\eeq
The warp factor $h$ is the solution of the following first-order differential equation:
\beq
h'\,+\,{2\over 3 r}\,h\,=\,-{512\over 792}\,{Q_c\over r^5}\,\,,
\eeq
whose general solution is:
\beq
h\,=\,{256\over 1215}\,{Q_c\over r^4}\,+\,{C\over r^{{2\over 3}}}\,\,,
\eeq
where $C$ is an integration constant.  In what follows we will take $C=0$. 

Let us write the resulting  total ten-dimensional metric as:
\beq
ds^2_{10}\,=\,ds^2_5\,+\,d\hat s_5^2\,\,,
\label{10d_flavored_metric}
\eeq
where $ds^2_5$ ($d\hat s_5^2$) is the metric of the non-compact (compact)  five-dimensional part of the ten-dimensional  space.  If we define the radius $R$ as:
\beq
R^4\,=\,{256\over 1215}\,Q_c\,\,,
\eeq
then $ds_5^2$ can be simply written as:
\beq
ds^2_5\,=\,{r^2\over R^2}\,\Big[dx^2_{1,2}\,+\,
\Big({4 Q_f\over 3}\Big)^{{4\over 3}}\,\,
{(dx^3)^2\over r^{{4\over 3}}}\Big]\,+\,
R^2\,{dr^2\over r^2}\,\,.
\label{anisotropic}
\eeq
This metric is an anisotropic version of $AdS_5$.  Let us  next write the compact part of the metric. First we define a new radius $\bar R$ as:
\beq
\bar R^2\,=\,{9\over 8}\,R^2\,\,.
\eeq
More explicitly, $\bar R^4$ is:
\beq
\bar R^4\,=\,{4\over 15}\,Q_c\,\,.
\eeq
Then, we can write  the compact metric $d\hat s_5^2$ as a squashed version of the original SE space:
\beq
d\hat s_5^2\,=\,\bar R^2\,\big[
\,ds^2_{KE}+{9\over 8}\,(d\tau+A)^2 \big]\,\,.
\label{compact_flavored_metric}
\eeq
Notice that the squashing factor in (\ref{compact_flavored_metric}) is constant and does not depend on the number of flavors.  Moreover, the non-compact part of the metric (written in (\ref{anisotropic})) is invariant under the following anisotropic scale transformations:
\beq
r\,\to\, r/\lambda\,\,,
\qquad\qquad
x^{0,1,2}\,\to \,\lambda\, x^{0,1,2}\,\,,
\qquad\qquad
x^{3}\,\to\,\lambda^{1\over 3}\,x^{3}\,\,,
\label{scaling_of_coordinates}
\eeq
where $\lambda$ is an arbitrary positive constant. This means that, effectively, the $x^3$ direction has an anomalous scale dimension.  According to the standard  notation, in a general Lifshitz-like anisotropic scaling the coordinates transform as in (\ref{scaling_of_coordinates}), with the anisotropic coordinate changing  as  $x^3\to \lambda^{1\over z}\,x^3$, where $z$ is an exponent which measures the degree of anisotropy of this coordinate. It is clear from (\ref{scaling_of_coordinates})  that $z=3$ for our system.  Notice also that the dilaton is not invariant under the scale transformation (\ref{scaling_of_coordinates}). Indeed, one can check easily that  it transforms as $e^{\phi}\to \lambda^{-{2\over 3}}\,e^{\phi}$. For other examples of scaling anisotropic backgrounds constructed from brane intersections see \cite{Azeyanagi:2009pr}. 

The existence of scale invariant solutions of our BPS equations is quite remarkable and it is consistent with the field theory analysis performed in \cite{Erdmenger:2002ex} of a bulk ${\cal N}=4$, $d=4$  theory coupled to a 
${\cal N}=4$, $d=3$ hypermultiplet living on a defect. Indeed, in this last reference it was shown that the model remains  superconformal after the addition of the hypermultiplet, giving rise to a defect conformal field theory. It would be very interesting to understand the value of the Lifshitz-like exponent $z$ in this context.

We have checked that the ten-dimensional metric (\ref{10d_flavored_metric})  is free of curvature singularities. The Ricci scalar and the squares of the Ricci  and Riemann tensors are constant and given by:
\beq
R^{\mu}_{\,\,\,\mu}\,=\,{7\sqrt{15}\over 4}\,Q_c^{-{1\over 2}}\,\,,
\qquad\,\,
R_{\mu\nu}\,R^{\mu\nu}\,=\,{3975\over 8}\,Q_c^{-1}\,\,,
\qquad\,\,
R_{\mu\nu\alpha\beta}\,R^{\mu\nu\alpha\beta}\,=\,{10095\over 16}\,Q_c^{-1}\,\,.
\eeq

Finally, let us point out that the Ricci scalar for a generic solution of the flavored BPS equations can be written in terms of $G$ and its radial derivative $G'=dG/dr$ as:
\beq
R^{\mu}_{\,\,\,\mu}\,=\,{Q_f\,e^{{3\phi\over 2}}\over h^{{1\over 2}}\,G}\,
\Big[\,{9\over G'}\,+\,{Q_f\,e^{{3\phi\over 2}}\over G}\,\Big]\,\,.
\label{R-general}
\eeq
Eq. (\ref{R-general})  shows that the solution is well-defined in the region in which $G$ is positive and monotonic ($R^{\mu}_{\,\,\,\mu}$ diverges if $G\not=0$ and $G'=0$).

\section{Massive flavors}
\label{massive}

In this section we discuss the generalization of the previous results to the situation in which the flavors added by the D5-branes are massive. In this case the flavor branes do not reach the origin at $r=0$ and one expects to have a smearing four-form $\Xi$ depending on the radial coordinate $r$. As noticed in \cite{Benini:2006hh} for other brane intersections, there is an easy way to incorporate this radial dependence of $\Xi$  by performing the following substitution in our ansatz:
\beq
Q_f\,\to\,Q_f\,p(r)\,\,,
\label{Q_f_massive}
\eeq
where $p(r)$ is a radial profile which depends on the particular  distribution of the flavor branes. We will assume that $p(r)$ is a monotonic function of $r$, which vanishes for $r$ smaller than certain fixed value $r=r_q$ and becomes equal to one in the deep UV region $r\to\infty$, where the quarks are effectively massless:
\beq
p(r<r_q)\,=\,0\,\,,
\qquad\qquad
p(r\to\infty)\,=\,1\,\,.
\eeq
Notice that now our problem has an explicit scale $r_q$, which corresponds to the mass of the quarks. In what follows we will assume that the function $p(r)$ is known. To obtain $p(r)$ for a given mass distribution of the flavors we would have to perform a microscopic calculation of the charge density of the D5-branes (see \cite{Bigazzi:2008zt} for a similar calculation for the D3-D7 system), something that we will not attempt to do here.  It follows from  this discussion that  the new ansatz for $F_3$ in this massive case  is:
\beq
F_3\,=\,Q_f\,p(r)\,dx^3\wedge {\rm Im }\,\hat\Omega_2\,\,\,.
\label{F3_ansatz_massive}
\eeq
The expression of the smearing four-form $\Xi$ can be immediately obtained by computing the exterior derivative of $F_3$ in (\ref{F3_ansatz_massive}).  We get:
\beq
2\,\kappa_{10}^2\,T_5\,\Xi\,=\,-Q_f\,\Big[\,3\,p(r)\,dx^3\wedge  {\rm Re }\,\hat\Omega_2\wedge (d\tau+A)\,+\,
p'(r)\,dx^3\wedge dr\wedge {\rm Im }\,\hat\Omega_2\,\big]\,\,.
\label{smearing_form_massive}
\eeq
Moreover, it is straightforward to check that the BPS equations are obtained from the ones in (\ref{BPS_system}) by performing the substitution (\ref{Q_f_massive}). It is also easy to verify that any solution of the BPS system also solves the equations of motion of the gravity-plus-branes system for any profile function $p(r)$. Indeed, $dF_5=0$ by construction and eqs. (\ref{eom_F3}), (\ref{eom_dilaton}) and (\ref{Einstein_eq}) are fulfilled if ${\cal K}_{(6)}$ is taken as in (\ref{K_6_expression}) and the energy-momentun tensor of the flavor brane is given by (\ref{Tmu-nu}), where the smearing form $\Xi$ is the one written in (\ref{smearing_form_massive}). 
In this massive case, the non-zero components of $T_{\m\n}^{\text{flav}}$, in flat coordinates, are given by:
\bear
&&T_{x^{\mu}\,x^{\nu}}^{\text{flav}}\,=\,-\,Q_f\,h^{-{1\over 2}}\,e^{{\phi\over 2}-f-2g-m}\,
\big( 3\,p\,+\,e^f\,p'\, \big)\eta_{\mu\nu}\,\,,
\qquad\qquad
(\mu,\nu=0,1,2)\,\,,\rc\rc
&&T_{rr}^{\text{flav}}\,=\,-\,3\,Q_f\,p\,h^{-{1\over 2}}\,e^{{\phi\over 2}-f-2g-m}\,\,,\rc\rc
&&T_{ij}^{\text{flav}}\,=\,-\,{Q_f\over 2}\,h^{-{1\over 2}}\,e^{{\phi\over 2}-f-2g-m}\,
\big( 3\,p\,+\,e^f\,p'\, \big)\delta_{ij}\,\,,
\qquad\qquad\,
(i,j=1,\cdots, 4)\,\,,\rc\rc
&& T_{55}\,=\,-Q_f\,h^{-{1\over 2}}\,e^{{\phi\over 2}-2g-m}\,p'\,\,.
\label{Tmu_nu_massive_components}
\eear

Let us now sketch the integration of the BPS system for any profile function $p(r)$. First of all, it is clear 
(\ref{phi_m}) continues to hold, \ie\ we can take $m=-\phi$ also in this massive case.  Moreover, the expression of $F_5$ can also be written in terms of $\phi$ and $h$ as in (\ref{F5_sol}).  Changing the radial coordinate as in (\ref{rho_coordinate_def}), we get the following equation for the dilaton:
\beq
\dot\phi\,=\,Q_f\,p\,e^{{3\phi\over 2}}\,\,,
\label{dot_phi_massive}
\eeq
where the dot denotes the derivative with respect to the new radial coordinate $\rho$. This  equation can be readily integrated, with the result:
\beq
e^{{3\phi\over 2}}\,=\,{
e^{{3\phi_q\over 2}}\over 1\,-\,{3\over 2}\,Q_f\,
e^{{3\phi_q\over 2}}\int_{\rho_q}^{\rho}\,p(\bar\rho)\,d\bar\rho}\,\,,
\label{dilaton_sol_massive}
\eeq
with $\phi_q\,=\,\phi(\rho=\rho_q)$ and $\rho_q$ is the value of $\rho$ corresponding to the  threshold value $r=r_q$ of the original $r$ variable, \ie\ $p(\rho<\rho_q)=0$. It is also straightforward to find a master equation generalizing (\ref{master_eq}).  The relation between $G$ and the functions $f$ and $g$ is the same as in 
(\ref{f_g_G}). Then,  $G$ satisfies the following non-linear second-order differential equation:
\beq
\ddot G\,-\,6\,G^2\,=\,{\dot \phi\over 2}\,\dot G\,\,.
\label{master_eq_massive}
\eeq
Notice that the right-hand side of (\ref{master_eq_massive})  vanishes for $\rho<\rho_q$ and 
 (\ref{master_eq_massive})  becomes, in this region,  identical to the unflavored equation (\ref{master_eq_unflavored}).  This is quite natural since there are no flavor sources when  $\rho<\rho_q$. On the contrary, when $\rho>\rho_q$ the dilaton runs according to (\ref{dilaton_sol_massive}) and the master equation gets deformed with respect to its unflavored version. Actually, it is quite convenient to write $\dot\phi$ in the following way. Let $\varepsilon_q$  be the flavor deformation parameter  at the threshold, namely:
\beq
\varepsilon_q\equiv {3\over 2}\,Q_f\,e^{{3\phi_q\over 2}}\,\,.
\label{epsilon_q}
\eeq
Then, by combining (\ref{dot_phi_massive}) and (\ref{dilaton_sol_massive}), we  get:
\beq
{\dot \phi\over 2}\,=\,{1\over 3}\,{\varepsilon_q\,p(\rho)\over
1\,-\,\varepsilon_q\,\int_{\rho_q}^{\rho}\,p(\bar\rho)\,d\bar\rho}\,\,.
\eeq
\begin{figure}[ht]
\center
 \includegraphics[width=0.50\textwidth]{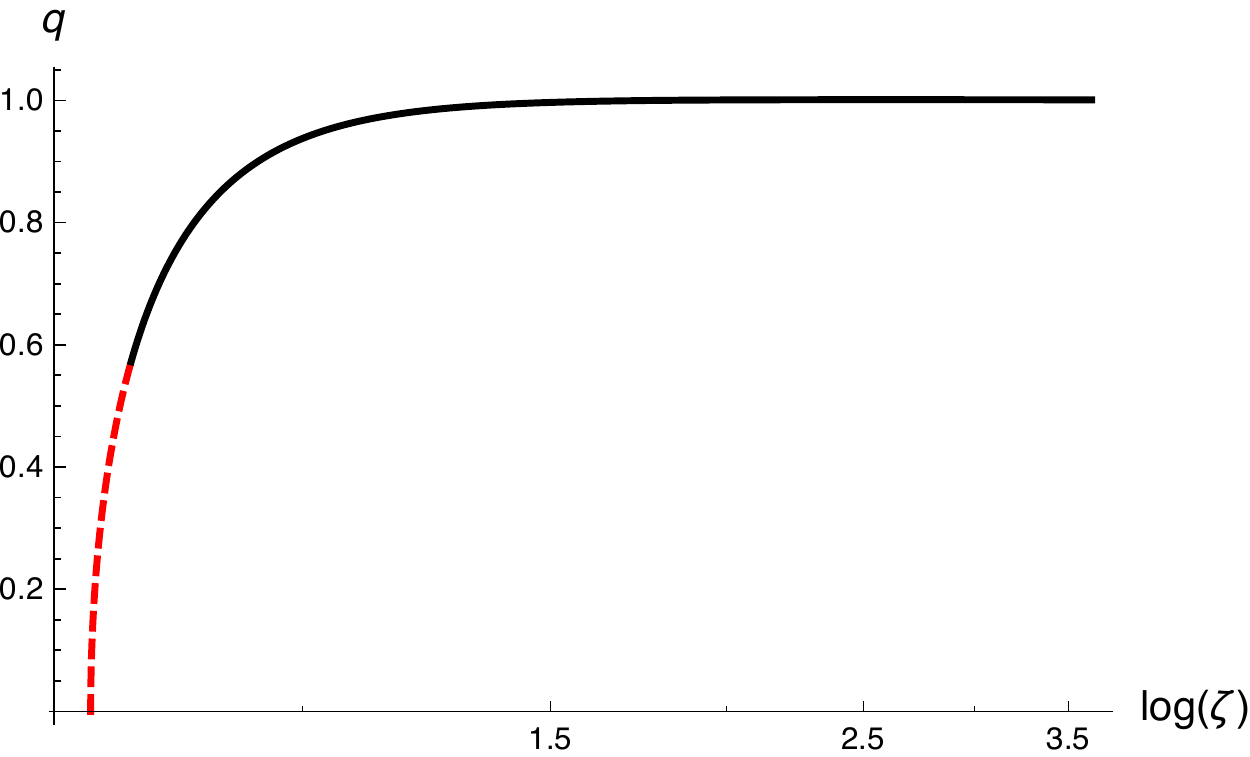}
  \caption{In this plot we depict the squashing function $q$ versus the holographic coordinate $\zeta=\sqrt{G}$
  for massive flavors with a Heaviside profile function $p$. 
  The unflavored curve  is the dashed line drawn in red. It corresponds to the Weierstrass  function  (\ref{general_unflavor_G}) with $g_3=0.5$. The flavor sources are placed at $\zeta\ge\zeta_q$, with $\zeta_q=0.755$. In this region the profile function $p$ is equal to one. The flavor deformation parameter for this plot is $\varepsilon_q=0.05$. The $q=q(\zeta)$ curve in this plot interpolates smoothly between $q=0$ and $q={3\over 2\sqrt{2}}\approx 1.06$.  }
\label{plot_squashing}
\end{figure}

Eq. (\ref{master_eq_massive}) must be solved numerically. Actually, we know analytically its solution in the region $\rho<\rho_q$,  where $G(\rho)$ is the Weierstrass function written on the right-hand side of (\ref{general_unflavor_G}).  We are interested in solutions of  (\ref{master_eq_massive})  approaching the scaling background described in section  \ref{flavored_sol} in the deep UV region $\rho\approx \varepsilon^{-1}$. To find this interpolating background we must solve numerically (\ref{master_eq_massive}) for $\rho>\rho_q$ by imposing the initial conditions at $\rho=\rho_q$ corresponding to the unflavored function (\ref{general_unflavor_G})  for some value of the lattice invariant $g_3$.  Next, we must find the value of $g_3$ which leads to the desired UV  behavior. This can be done by means of the standard shooting technique. A good strategy to perform this calculation is by considering the relative squashing of the internal manifold, defined as:
 \beq
 q\equiv e^{f-g}\,\,.
 \label{q_def}
 \eeq
 From (\ref{f_g_G}) we obtain that $q$ is related to our master function $G$ as:
  \beq
 q\,=\,{\dot G\over 2 G^{{3\over 2}}}\,\,.
 \eeq
The flavored solution of section \ref{flavored_sol} has a constant squashing given by $q={3\over 2\sqrt{2}}$, while for the unflavored general solution of section \ref{unflavored_sol}  $q=\sqrt{k}$, \ie\ $q\to 0 (1)$ in the IR (UV) if the parameter $b\not=0$ and  it takes the constant value $q=1$ if $b=0$. For a given profile function $p$ we want to integrate (\ref{master_eq_massive}) in such a way that $q$ interpolates between the value $q=0$ in the IR and $q={3\over 2\sqrt{2}}$ in the UV. We have verified numerically that this is possible for the simplified case in which $p(\rho)$ is a Heaviside step function ($p(\rho)=\Theta(\rho-\rho_q)$). This fact is illustrated in Figure \ref{plot_squashing} for some particular values of the parameters.

\section{Summary and outlook}
\label{conclusions}

In this paper we found supersymmetric solutions generated by the intersection of color D3-branes and flavor D5-branes, in which the latter create a codimension one defect on the worldvolume of the color branes. Our backgrounds solve the equations of motion of supergravity with sources corresponding to smeared flavor branes. We considered a generic case in which the color D3-branes are placed on the tip of a Calabi-Yau cone 
constructed with a general Sasaki-Einstein space.  We found a system of first-order BPS equations, which we were able to integrate in general in the unflavored case. Moreover, we obtained a particular solution of the flavored equations for massless quarks which gives rise to a metric displaying anisotropic scale invariance.
Finally, we extended our ansatz to the case of massive flavors. 

In this work we restricted ourselves to finding the unquenched geometry corresponding to the D3-D5 system. A natural extension of our results would be the calculation of different observables and the comparison with the ones obtained with the probe approximation. We could, for example, compute the meson spectrum and compare the results with the ones found in \cite{Arean:2006pk}. In order to perform this calculation we must add an additional flavor D5-brane, treated as a probe,  and study its fluctuations in the unquenched background. 
In the case of massive flavors we only sketched the form of the solutions. What remains to be done is the calculation of the profile $p(r)$ for a given quark mass and the integration of the master equation. Once this is done we could study the dependence of the observables on the quark mass. When the mass of the quarks is increased the flavor effects should decrease and, in fact, they should disappear completely when the quarks are infinitely massive. Therefore, by varying the quark mass we introduce a renormalization group flow and we could analyze, for example, how the entanglement entropy changes with the energy scale or how the anisotropy of the system evolves with the quark mass.

The backgrounds constructed here could be generalized in several directions. The most obvious one would be the construction of a black hole solution for the D3-D5 system, which would be the dual of the unquenched defect theory at non-zero temperature (see \cite{Bigazzi:2009bk} for a similar problem in the D3-D7 system). 
We could even add charge density and magnetic field and try to determine the phase diagram in the Veneziano limit.  Another interesting direction to follow would be the generalization of the conifold case to include cascading theories of the Klebanov-Strassler type.  We could also compactify the $x^3$ coordinate and impose antiperiodic  boundary conditions to the fermions. This would break supersymmetry and would give rise to a cigar-shaped geometry dual to a confining theory in three spacetime directions.  

Another interesting research project  is the study of the phenomenon of magnetic catalysis in $2+1$ dimensions. The presence of a magnetic field drives spontaneous chiral symmetry breaking even when the field strength is weak. Unquenching the magnetic catalysis requires the smearing of massive flavors in the presence of an external magnetic field (see \cite{Erdmenger:2011bw, Filev:2011mt} for the calculation in $3+1$ dimensions). Notice also that our solutions are smeared along the $x^3$ direction. By performing a T-duality along $x^3$ we would get a backreacted D2-D6 solution in the type IIA theory, whose transverse space contains a flat direction along which the D2-branes are smeared.  It would be very interesting to compare this solution with other D2-D6 backreacted backgrounds.

We are currently addressing some of these problems and we hope to report on them in the near future.

\vspace{1cm}
{\bf \large Acknowledgments}

\noindent 
We are grateful to F. Bigazzi, V. Filev, G. Itsios, N. Jokela, D. Mateos, C. N\'u\~nez,  A. Paredes and D. Rosa for discussions and useful suggestions.  The work of E. C.  was supported in part by the National Research Foundation of Korea through the grant NRF-2014R1A6A3A04056670, and the grants 2005-0093843, 2010-220-C00003 and 2012K2A1A9055280.
H. L. is supported in part by NSF grants DMS-1159412, PHY-0937443 and PHY-1306313, and in part by YMSC, Tsinghua University. J. M. P. and A. V. R.   are funded by the Spanish grant FPA2014-52218-P by Xunta de Galicia (GRC2013-024), and by FEDER. J. M.  P. is supported by the Spanish FPU fellowship FPU14/06300.
D. Z. acknowledges financial support from the Universitat Internacional de Catalunya. Centro de F\'{i}sica do Porto is partially funded by FCT through the project CERN/FIS-NUC/0045/2015.

\appendix

\vskip 3cm
\renewcommand{\theequation}{\rm{A}.\arabic{equation}}
\setcounter{equation}{0}
%\medskip

\section{Supersymmetry analysis}

In this appendix we analyze the Killing spinor equations for our ansatz. We will begin by writing the supersymmetry variations of the dilatino $\lambda$ and gravitino $\psi$  in a background of type IIB supergravity with RR three- and five-forms. In the Einstein frame, these variations are:
\bear
&& \d\l\,=\,\frac{1}{2}\Gamma^{a}\partial_a \f-\frac{e^{\f/2}}{24}(F_3)_{abc}\Gamma^{abc}\s_1\eps\,,\rc\rc
&& \d\j_{a}\,=\,D_a\epsilon\,+\,{\frac{1}{4}\frac{1}{480}}\, (F_5)_{bcdef}\Gamma^{bcdef}\Gamma_a i\s_2\eps \,
 -{\frac{1}{96}}e^{\f/2}(F_3)_{bcd}\left({\Gamma_{a}}^{bcd}-9{\d_{a}}^{b}\Gamma^{cd}\right)\s_1\eps\,,
 \qquad\qquad
 \label{SUSY_variations}
  \eear
 where $\sigma_1$ and $\sigma_2$ are Pauli matrices and $\epsilon$ is a doublet of Majorana-Weyl spinors. In 
 (\ref{SUSY_variations})  $D_a\epsilon=(\partial_a+\frac{1}{4}{\o_{a}}^{bc}\Gamma_{bc})\,\eps$ is the covariant derivative acting on the Killing spinor $\epsilon$.  We will work in flat components with respect to the following vielbein basis:
\bear
&&E^{x^{\mu}}\,=\,h^{-{1\over 4}}\,dx^{\mu}\,\,,\qquad\qquad (\mu=0, 1,2),
\qquad\qquad
E^{x^{3}}\,=\,h^{-{1\over 4}}\,e^{m}\,dx^{3}\,\,,\rc\rc
&&E^{r}\,=\,h^{{1\over 4}}\,dr\,\,,\rc\rc
&&E^{i}\,=\,h^{{1\over 4}}\,e^g\,e^i\,\,,
\qquad\qquad (i=1,\cdots, 4)\,\, ,
\qquad\qquad
E^{5}\,=\,h^{{1\over 4}}\,e^f\,(d\tau+A)\,\,.
\label{vielbein}
\eear
Let us write the RR forms in flat components. The five-form (\ref{F5_ansatz}) can be written as: 
\beq
F_5\,=\,h^{{3\over 4}}\,e^{-m}\,K\,(1+*)\,
E^{x^0}\wedge E^{x^1}\wedge E^{x^2}\wedge E^{x^3}\wedge E^{r}
\eeq
Using (\ref{eom_F5}),  this last expression becomes:
\beq
F_5\,=\,Q_c\,h^{-{5\over 4}}\,e^{-4g-f}\,(1+*)\,
E^{x^0}\wedge E^{x^1}\wedge E^{x^2}\wedge E^{x^3}\wedge E^{r}\,\,.
\eeq
Moreover, the three-form $F_3$ written in (\ref{F3_ansatz_explicit}) can be recast as:
\beq
F_3\,=\,Q_f\,h^{-{1\over 4}}\,e^{-m-2g}\,
E^{x^3}\wedge \Big[E^1\wedge (\cos (3\tau)\,E^4+\sin(3\tau)\, E^3\,)\,+\,
E^2\wedge (\cos (3\tau)\,E^3-\sin(3\tau)\, E^4\,)\Big]\,\,.
\eeq

Let us start our analysis by considering  the condition $\delta \lambda=0$. First we impose to the spinor $\epsilon$  the projection required by the   K\"ahler condition:
\beq
\Gamma^{12}\,\epsilon\,=\,\Gamma^{34}\,\epsilon\,\,.
\eeq
Then, it is straightforward to show that $\delta\lambda=0$ implies:
\beq
\phi'\,\epsilon\,-\,Q_f\,e^{{\phi\over 2}\,-\,m\,-2g}\,
\Big[\cos (3\tau)\,\Gamma^{14}\,-\,\sin(3\tau)\,\Gamma^{24}\,\Big]\,\Gamma^{r\,x^3}\,\sigma_1\,\epsilon\,=\,0\,\,.
\eeq
Therefore, for consistency, we  must impose the following additional projection to $\epsilon$:
\beq
\Big[\cos (3\tau)\,\Gamma^{14}\,-\,\sin(3\tau)\,\Gamma^{24}\,\Big]\,\Gamma^{r\,x^3}\,\sigma_1\,\epsilon\,=\,\epsilon\,\,,\label{dilatino_proj}
\eeq
and  we get the following differential equation for the dilaton:
\beq
\phi'\,=\,Q_f\,e^{{\phi\over 2}\,-\,m\,-2g}\,\,.
\label{phi_prime_eq}
\eeq
We go on by requiring that $\delta\psi_a=0$ for the different directions of the ten-dimensional spacetime. Due to the presence of the covariant derivative of the spinor, we need the different components of the spin connection one-form.  The non-vanishing components are   easily computed for our vielbein basis by using Cartan's structure equation,  with the result:
\bear
&&\omega^{x^{\mu}}_{\,\,\,\,\,r}\,=\,-{1\over 4}\,h'\,h^{-{5\over 4}}\,E^{x^{\mu}}\,\,,
\qquad\qquad
(\mu=0,1,2)\,\,,\rc\rc
&&\omega^{x^{3}}_{\,\,\,\,\,r} \,=\,\Big(-{1\over 4}\,h'\,h^{-{5\over 4}}+m'\,h^{-{1\over 4}}\Big)\,E^{x^3}\,,\rc\rc
&&\omega^{i }_{\,\,\,\,\,r}\,=\,\Big({1\over 4}{h'\over h}\,+\,g'\Big)\,h^{-{1\over 4}}\,E^i\,\,,
\qquad\qquad
(i=1,\cdots, 4)\,\,,\rc\rc
&&\omega^{5}_{\,\,\,\,\,r}\,=\,\Big({1\over 4}{h'\over h}\,+\,f'\Big)\,h^{-{1\over 4}}\,E^5\,\,,\rc\rc
&&\omega^{5}_{\,\,\,\,\,i}\,=\,e^{f-2g}\,h^{-{1\over 4}}\,J_{ij}\,E^j\,\,,
\qquad\qquad(i=1,\cdots, 4)\,\,,\rc\rc
&&\omega^{i}_{\,\,\,\,\,j}\,=\,\hat \omega^{i}_{\,\,\,\,\,j}\,-\,e^{f-2g}\,h^{-{1\over 4}}\,J^{i}_{\,\,\,\,\,j}\,E^{5}\,\,,
\eear
where $\hat \omega^{i}_{\,\,\,\,\,j}$ is the spin connection one-form of the KE base, which satisfies
$de^i+\hat\omega^{i}_{\,\,\,\,\,j}\,\wedge\,e^j\,=\,0$.

Let us next require that $\delta \psi_{x^{\mu}}=0$, for $\mu=0,1,2$. We get:
\bear
&&h^{-{5\over 4}}\,h'\,\Gamma_{x^{\mu}\,r}\,\epsilon\,-\,{Q_c\over 2}\,e^{-4g-f}\,h^{-{5\over 4}}\,
\Big(\Gamma^{x^0\,x^1\,x^2\,x^3\,r}\,+\,\Gamma^{12345}\Big)\,\Gamma_{x^{\mu}}\,i\sigma_2\,\epsilon\,+\,\rc\rc
&&\qquad\qquad-
 Q_f\, e^{{\phi\over 2}\,-\,m\,-2g}\,h^{-{1\over 4}}\,
 \Big[\cos (3\tau)\,\Gamma^{14}\,-\,\sin(3\tau)\,\Gamma^{24}\,\Big]\,\Gamma^{x^{\mu}\,x^3}\,\sigma_1\,\epsilon\,=\,0\,\,,
 \qquad
 \label{gravitino_012}
\eear
We now take into account the ten-dimensional chirality condition satisfied by the spinor $\epsilon$:
\beq
\Gamma^{x^0\,x^1\,x^2\,x^3\,r\,1\,2\,3\,4\,5}\epsilon\,=\,\epsilon\,\,,
\label{total_chirality}
\eeq
which implies that:
\beq
\Gamma^{x^0\,x^1\,x^2\,x^3\,r}\epsilon\,=\,-
\Gamma^{1\,2\,3\,4\,5}\epsilon\,\,,
\label{total_chirality_2}
\eeq
and we impose the following projection:
\beq
\Gamma_{x^0\,x^1\,x^2\,x^3}\,(i\sigma_2)\,\epsilon\,=\,\epsilon\,\,,
\label{D3-brane-proj}
\eeq
which corresponds to having a D3-brane extended in the $x^0\,x^1\,x^2\,x^3$ directions.  Using (\ref{total_chirality_2}), (\ref{D3-brane-proj})  and (\ref{dilatino_proj})
we conclude  that (\ref{gravitino_012}) implies the following differential equation for the warp factor $h$:
\beq
h'\,=\,-Q_c\,e^{-4g-f}\,-\,Q_f\,e^{{\phi\over 2}-m-2g}\,h\,\,.
\label{hprime_eq}
\eeq
We now consider the supersymmetric  variation of the gravitino along the $x^3$ direction. We get:
\bear
&&\Big(h^{-{5\over 4}}\,h'\,-\,4\,h^{-{1\over 4}}\,m'\Big)\,\Gamma_{x^3 r}\,\epsilon\,-\,
{Q_c\over 2}\,e^{-4g-f}\,h^{-{5\over 4}}\,
\Big(\Gamma^{x^0\,x^1\,x^2\,x^3\,r}\,+\,\Gamma^{12345}\Big)\,\Gamma_{x^{3}}\,(i\sigma_2)\,\epsilon\,-\,\rc\rc
&&\qquad\qquad
-3\,Q_f\,e^{{\phi\over 2}\,-\,m\,-2g}\,h^{-{1\over 4}}\,
 \Big[\cos (3\tau)\,\Gamma^{14}\,-\,\sin(3\tau)\,\Gamma^{24}\,\Big]\,\sigma_1\,\epsilon\,=\,0\,\,.
\eear
Using (\ref{total_chirality_2}), (\ref{D3-brane-proj}) and (\ref{dilatino_proj}), we get the following differential equation for $m$:
\beq
m'\,=\,{h'\over 4h}\,+\,{Q_c\over 4}\,e^{-4g-f}\,h^{-1}\,-\,{3\over 4}\,Q_f\,e^{{\phi\over 2}\,-\,m\,-2g}\,\,.
\eeq
Plugging (\ref{hprime_eq}) on the right-hand side of this last equation, we get
\beq
m'\,=-\,Q_f\,e^{{\phi\over 2}\,-\,m\,-2g}\,\,.
\label{m_prime_eq}
\eeq
By comparing with (\ref{phi_prime_eq}) we conclude that $m'=-\phi'$. 

Let us next consider the variation of the gravitino along the radial direction, which leads to the following equation:
\bear
&&h^{-{1\over 4}}\,\partial_r\,\epsilon+{Q_c\over 16}\, e^{-4g-f}\,h^{-{5\over 4}}\, \Gamma^r\,
\Big(\Gamma^{x^0\,x^1\,x^2\,x^3\,r}\,-\,\Gamma^{12345}\Big)\,i\sigma_2\,\epsilon\,-\,\rc\rc
&&\qquad\qquad-
{Q_f\over 8}e^{{\phi\over 2}\,-\,m\,-2g}\,h^{-{1\over 4}}\,\Gamma^{r\,x^3}\,
 \Big[\cos (3\tau)\,\Gamma^{14}\,-\,\sin(3\tau)\,\Gamma^{24}\,\Big]\,
 \sigma_1\,\epsilon\,=\,0\,\,.
\eear
After imposing again (\ref{dilatino_proj}), (\ref{total_chirality_2}) and (\ref{D3-brane-proj}), this equation becomes:
\beq
\partial_r\,\epsilon\,=\,\Big(
{Q_c\over 8}\, e^{-4g-f}\,h^{-1}\,+\,{Q_f\over 8}\,e^{{\phi\over 2}\,-\,m\,-2g}\Big)\epsilon\,\,.
\label{radial_gravitino_proyected}
\eeq
Comparing with (\ref{hprime_eq}), the right-hand side of (\ref{radial_gravitino_proyected}) can be written in terms of the derivative of the warp factor $h$:
\beq
\partial_r\,\epsilon\,=\,-{h'\over 8 h}\,\epsilon\,\,.
\label{radial_derivative_epsilon}
\eeq
This equation can be integrated as:
\beq
\epsilon\,=\,h^{-{1\over 8}}\,\tilde \epsilon\,\,,
\label{radial_dependence_epsilon}
\eeq
with $\tilde \epsilon$ being a spinor  independent of the radial coordinate. 

Let us next analyze the variations of the components of the gravitino along the internal SE space.  We first consider one of the KE directions (say the  $e^1$ direction).  We get:
\bear
&&e^{-g}\,\Gamma_{1r}\,\Big(\hat D_1\,\epsilon-A_1\partial_{\tau}\epsilon\Big)\,+\,
{1\over 2}\,e^{f-2g}\,\Gamma_{5r12}\,\epsilon\,-\,{1\over 2}\,\Big({h'\over 4h}+g'\Big)\,\epsilon\,-\,\rc\rc
&&\qquad\qquad\qquad\qquad-
{Q_c\over 8}\,e^{-4g-f}\,h^{-1}\,\epsilon\,-\,{Q_f\over 8}\,e^{{\phi\over 2}-m-2g}\,\epsilon\,=\,0\,\,,
\label{gravitino_KE}
\eear
where $\hat D$ is the spinor  covariant derivative on the KE base (\ie\ with the spin connection $\hat \omega^{ij}$) and $A$ is the one-form potential of $J_{KE}$ (see (\ref{KE_pot})).  From (\ref{gravitino_KE}) it is clear that we must impose a further projection on $\epsilon$, namely:
\beq
\Gamma_{5r12}\,\epsilon\,=\,\epsilon\,\,.
\label{Gamma_5r12_projection}
\eeq
One can check that this new projection combined with the ones previously imposed to $\epsilon$ implies:
\beq
\Gamma_{12}\,\epsilon\,=\,\Gamma_{34}\,\epsilon\,=\,\Gamma_{r5}\,\epsilon\,=\,i\sigma_2\,\epsilon\,\,.
\eeq
We now use the fact that in any KE space there is a covariantly constant spinor which satisfies the condition:
\beq
\hat D_i\,\epsilon\,=\,{3\over 2}\,\Gamma_{12}\,A_i\,\epsilon\,=\,{3\over 2}\,(i\sigma_2)\,A_i\,\epsilon\,\,.
\label{KE_cov_constant}
\eeq
Actually, in the vielbein basis  of the  KE space   we are using it turns out that $\epsilon$ can be taken to be independent of the KE coordinates and (\ref{KE_cov_constant}) follows from our projections. Moreover, the difference appearing on the first term in (\ref{gravitino_KE}) (for any  KE direction) becomes:
\beq
\hat D_i\,\epsilon\,-\,A_i\,\partial_{\tau}\,\epsilon\,=\,A_i\,\Big({3\over 2}\,(i\sigma_2)\,\epsilon\,-\,\partial_{\tau}\,\epsilon\Big)\,\,,
\eeq
and clearly vanishes if we  require that $\epsilon$ depends on $\tau$ in such a way that:
\beq
\partial_{\tau}\,\epsilon\,=\,{3\over 2}\,(i\sigma_2)\,\epsilon\,=\,{3\over 2}\,\Gamma_{12}\,\epsilon\,\,.
\label{tau_derivative_epsilon}
\eeq
Using (\ref{Gamma_5r12_projection}) and (\ref{tau_derivative_epsilon}),  we arrive at the following differential equation for the function $g$:
\beq
g'\,=\,-{h'\over 4h}\,+\,e^{f-2g}\,-\,{Q_c\over 4}\,e^{-4g-f}\,h^{-1}\,-\,{Q_f\over 4}\,e^{{\phi\over 2}-m-2g}\,\,.
\label{gprime_hprime}
\eeq
Let us now substitute the value (\ref{hprime_eq}) of $h'$ on the right-hand side of (\ref{gprime_hprime}). We get the following equation for $g$:
\beq
g'\,=\,e^{f-2g}\,\,.
\label{g_prime_eq}
\eeq
Let us finally consider the variation of the gravitino along the SE fiber $\tau$. We get:
\beq
\Gamma_{5r}\Big(e^{-f}\,\partial_{\tau}\epsilon-e^{f-2g}\,\Gamma_{12}\,\epsilon\Big)-
{1\over 2}\,\Big({h'\over 4h}\,+\,f'\Big)\,\epsilon-
{Q_c\over 8}\,e^{-4g-f}\,h^{-1}\,\epsilon+{Q_f\over 8}\,e^{{\phi\over 2}-m-2g}\,\epsilon=0\,\,,
\eeq
After using (\ref{Gamma_5r12_projection}) and (\ref{tau_derivative_epsilon}), we arrive at the following first-order equation for $f$:
\beq
f'\,=\,-{h'\over 4h}\,+\,3e^{-f}\,-\,2e^{f-2g}\,-\,{Q_c\over 4}\,e^{-4g-f}\,h^{-1}\,+\,{Q_f\over 4}\,e^{{\phi\over 2}-m-2g}\,\,.
\eeq
which, after using (\ref{hprime_eq}), becomes:
\beq
f'\,=\,3\,e^{-f}\,-\,2e^{f-2g}\,+\,{Q_f\over 2}\,e^{{\phi\over 2}-m-2g}\,\,.
\label{f_prime_eq}
\eeq
Collecting (\ref{phi_prime_eq}), (\ref{hprime_eq}), (\ref{m_prime_eq}), (\ref{g_prime_eq}) and (\ref{f_prime_eq}),  we obtain the BPS first-order system (\ref{BPS_system}). 

Let us finally find the expression of the spinor $\epsilon$ satisfying all the conditions we have imposed. We first notice that (\ref{dilatino_proj}) can be written as:
\beq
\Gamma^{rx^314}\,e^{-3\tau\,\Gamma_{12}}\,\sigma_1\,\epsilon\,=\,\epsilon\,\,,
\eeq
and is solved by a spinor of the form:
\beq
\epsilon\,=\,h^{-{1\over 8}}\,e^{{3\over 2}\,\Gamma_{12}\,\tau}\,\,\eta\,\,,
\label{spinor_sol}
\eeq
where $\eta$ is a constant spinor which satisfies the following projection equation:
\beq
\Gamma_{rx^314}\,\sigma_1\,\eta\,=\,\eta\,\,.
\label{D5_projection}
\eeq
In (\ref{spinor_sol}) we have already taken into account the dependence of $\epsilon$ on the radial coordinate 
written in (\ref{radial_dependence_epsilon}). It is now immediate to check that all the  conditions required are fulfilled   if $\eta$ is constant and, besides (\ref{D5_projection}), satisfies the equation:
\beq
\Gamma_{12}\,\eta\,=\,\Gamma_{34}\,\eta\,=\,\Gamma_{r5}\,\eta\,=\,i\sigma_2\,\eta\,\,.
\label{CY_proyection}
\eeq
Notice that, after using (\ref{CY_proyection}), the expression of $\epsilon$ written in (\ref{spinor_sol})  is the same as in (\ref{Killing_spinor}). 
\subsection{Some particular cases}
Let us finish this appendix by writing the coordinate representation of the metric of two particularly relevant Sasaki-Einstein spaces. First of all we consider the case in which ${\cal M}_5=T^{1,1}$, whose KE base is 
the product ${\mathbb S}^2\times {\mathbb S}^2$.  The vielbein one-forms $e^i$ can be parameterized as:
\bear
&&e^1\,=\,{1\over \sqrt{6}}\,\sin\theta_1\,d\phi_1\,\,,
\qquad\qquad
e^2\,=\,{1\over \sqrt{6}}\,d\theta_1\,\,,\rc\rc
&&e^3\,=\,{1\over \sqrt{6}}\,\sin\theta_2\,d\phi_2\,\,,
\qquad\qquad
e^4\,=\,{1\over \sqrt{6}}\,d\theta_2\,\,,
\eear
where $\theta_i$ and $\phi_i$ are  angles which take values in the range $0\le\theta_1, \theta_2\le \pi$ and $0\le\phi_1, \phi_2<2 \pi$.  The fiber coordinate $\tau$ in the $T^{1,1}$  space  is usually represented as
$\tau=\psi/3$, where $0\le \psi<4\pi$, and the one-form $A$ takes the form:
\beq
A\,=\,{1\over 3}\big(\cos\theta_1\,d\phi_1\,+\,\cos\theta_2\,d\phi_2\big)\,\,.
\eeq
From this coordinate representation it is straightforward to compute the volume of the 
$T^{1,1}$ space, as well as the  value of corresponding constant $Q_c$. We get:
\beq
{\rm Vol}\big(T^{1,1}\big)\,=\,{16\over 27}\,\,\pi^3\,\,,
\qquad\qquad
Q_c\big(T^{1,1}\big)\,=\,27\,\pi\,g_s\,\alpha'{}^{\,2}\,\,N_c\,\,.
\eeq

Our second example is the five-sphere ${\mathbb S}^5$, which is a SE space with ${\mathbb C}\,{\mathbb P}^2$ base. 
In order to represent the four-dimensional metric of the  ${\mathbb C}\,{\mathbb P}^2$ base, let us consider  an angular coordinate $\chi$ taking values in the range $0\le \chi \le \pi$, as well as a set of $SU(2)$  left-invariant  one-forms $\omega^i$ ($i=1,2,3$) satisfying $d\omega^i\,=\,{1\over 2}\,\epsilon^{ijk}\,\omega^j\wedge \omega^k$.  Then, the vielbein basis of ${\mathbb C}\,{\mathbb P}^2$ is:
\bear
&&e^1\,=\,{1\over 2}\cos\big({\chi\over 2})\,\omega^1\,\,,
\qquad\qquad\qquad\,\,\,\,\,\,\,
e^2\,=\,{1\over 2}\cos\big({\chi\over 2})\,\omega^2\,\,,\rc\rc
&&e^3\,=\,{1\over 2}\cos\big({\chi\over 2})\,\sin\big({\chi\over 2})\,\omega^3\,\,,
\qquad\qquad
e^4\,=\,{1\over 2}\,d\chi\,\,.
\eear
In this case the fiber $\tau$ takes values in the range $0\le\tau\le 2\pi$ and the one-form $A$  is:
\beq
A\,=\,{1\over 2}\,\cos^2\big({\chi\over 2}\big)\,\omega^3\,\,.
\eeq
Finally, after a simple calculation,  we readily  obtain:
\beq
{\rm Vol}\big({\mathbb S}^5\big)\,=\,\pi^3\,\,,
\qquad\qquad
Q_c\big({\mathbb S}^5\big)\,=\,16\,\pi\,g_s\,\alpha'{}^{\,2}\,\,N_c\,\,.
\eeq

\end{document}